\documentclass[a4paper,notitlepage,tightenlines,twocolumn,nofootinbib,superscriptaddress,preprintnumbers,showpacs]{revtex4-1}

\usepackage{times}
\usepackage{mathptmx}   

\usepackage[colorlinks,citecolor=blue,linktoc=all,linkcolor=red]{hyperref}
\usepackage{amssymb,amsbsy,amsmath,amsfonts}
\usepackage{mathrsfs} 
\usepackage{nicefrac}
\usepackage{upgreek}
\usepackage{esint}
\usepackage{slashed}

\usepackage{graphicx}
\usepackage{graphics,psfrag}

\DeclareMathAlphabet{\mathantt}{OML}{antt}{l}{it}
\DeclareMathAlphabet{\mathpzc}{OT1}{pzc}{m}{n}

\def\beq{\begin{equation}}
\def\eeq{\end{equation}}
\def\bea{\begin{eqnarray}}
\def\eea{\end{eqnarray}}
\def\beqa{\begin{equation}\begin{array}{l}}
\def\eeqa{\end{array}\end{equation}}
\def\eqlab#1{\label{eq:#1}}
\def\figlab#1{\label{fig:#1}}

\def\seclab#1{\label{sec:#1}}

\def\Eqref#1{Eq.~(\ref{eq:#1})}

\def\Figref#1{Fig.~\ref{fig:#1}}

\def\secref#1{Sec.~\ref{sec:#1}}

\def\half{\mbox{$\frac{1}{2}$}}

\def\barr{\left(\begin{array}{c}}
\def\earr{\end{array}\right)}
\def\bmat{\left(\begin{array}{cc}}
\def\emat{\end{array}\right)}
\def\al{\alpha}
\def\be{\beta}
\def\ga{\gamma} 
\def\de{\delta}

\def\la{\lambda}

\def\si{\sigma}

\def\nn{\nonumber}
\def\nn{\nonumber}
\def\dd{\mathrm{d}}

\DeclareMathOperator\im{Im}

\def\3d{3-D}

\begin{document}

\title{Dissecting the
hadronic contributions to $(g-2)_\upmu $ 
by Schwinger's sum rule}
\author{Franziska Hagelstein}
\affiliation{
Institut f\"ur Kernphysik \& Cluster of Excellence PRISMA,  Johannes Gutenberg-Universit\"at Mainz, D-55128 Mainz, Germany}
\affiliation{Albert Einstein Center for Fundamental Physics, Institute for Theoretical Physics,\\ University of Bern,  CH-3012 Bern, Switzerland}
\author{Vladimir Pascalutsa}
\affiliation{
Institut f\"ur Kernphysik \& Cluster of Excellence PRISMA,  Johannes Gutenberg-Universit\"at Mainz, D-55128 Mainz, Germany}

\begin{abstract}
The theoretical uncertainty of 
$(g-2)_\upmu $ is currently dominated by hadronic contributions.    In order to express those in terms
of directly measurable quantities, we consider a sum rule
relating $g-2$ to  an integral of
a photo-absorption cross section.  The sum rule, attributed to 
Schwinger, can be viewed as a combination of two older
sum rules:
Gerasimov-Drell-Hearn and Burkhardt-Cottingham. The Schwinger
sum rule has an important feature, distinguishing it from the other two: 
the relation between the anomalous magnetic moment and
the integral of a photo-absorption cross section
is linear, rather than quadratic. The linear property makes 
it suitable for a straightforward assessment of the hadronic 
contributions to  $(g-2)_\upmu $. From the sum rule we
rederive the Schwinger $\alpha/2\pi$ correction, as well as the
formula for the hadronic vacuum-polarization contribution. 
As an example of the light-by-light contribution we consider 
the single-meson exchange. 
\end{abstract}
\pacs{13.60.-r, 11.55.Hx, 14.60.-z, 14.80.Va}
\preprint{MITP/17-042}

\maketitle
\thispagestyle{empty}

\section{Introduction}
The anomalous magnetic moment (AMM) of the muon, $\varkappa_\upmu\equiv \nicefrac12 (g-2)_\upmu$,
serves as a stringent precision 
tests of the Standard Model (SM).  And at present  it
does not work out for the SM --- the
experimental value is about $3\sigma$ away from the SM
prediction \cite{Blum:2013xva,Benayoun:2014tra}.
While the uncertainties of the SM and the experimental value are comparable,
the new Fermilab experiment \cite{Logashenko:2015xab,Venanzoni:2014ixa} will, in a few years, 
reduce the experimental error-bar by nearly a factor of four.  
The prospects
for reducing the SM (theory) uncertainty are, on the other hand,
more obscure.  The SM error-bar is dominated by the
{\it hadronic contributions},  which are very difficult to compute in the SM
due to  the non-perturbative nature of Quantum Chromodynamics
(QCD).
In the present SM value these contributions  are determined
empirically, using general relations to other
experimental observables in combination with mesonic model calculations, rather
than from QCD directly. 
It is the necessity of resorting 
to models --- particularly in evaluation of the so-called
hadronic light-by-light (HLbL) contributions [c.f.\ \Figref{Hamm}(b)] --- 
which makes it difficult
to reduce the uncertainty of the current SM value.

In the future, lattice QCD  
will deliver a sufficiently precise 
{\it ab initio} calculation of the HLbL contribution;
for recent progress in this direction see Refs.~\cite{Blum:2014oka,Blum:2015gfa,Green:2015sra,Green:2015mva}. 
Until then, the best hope for improvement 
is to replace the model evaluations with 
model-independent, ``data-driven''  approaches
based on dispersion theory.
The data-driven approach is fairly well-founded
and routinely used for the hadronic vacuum-polarization (HVP) contribution [\Figref{Hamm}(a)], since it can exactly be written as a dispersion integral of the decay rate of a virtual timelike
photon into hadrons, which to a good approximation is expressed in terms of the observed ratio $\mu^+\mu^-/e^+e^-\to \mathrm{hadrons}$, see
e.g., Refs.~\cite{Jegerlehner:2017gek,Davier:2016iru}.
The HLbL contribution is much more complicated from this point of view,
because it involves the dispersion relations for 3- and/or 4-point functions,
rather than for a 2-point function as in case of HVP, see 
Refs.~\cite{Colangelo:2014pva,Colangelo:2014dfa,Colangelo:2017qdm,Colangelo:2017fiz} and \cite{Pauk:2014rfa} for the two recent approaches to this problem.

\begin{figure}[b]
\includegraphics[width=0.32\textwidth]{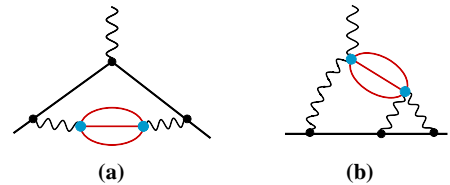}
\caption{Hadronic contributions to $(g-2)_\upmu$: (a) HVP, (b) HLbL.}
\figlab{Hamm}
\end{figure}

Here we consider yet another  
approach to a data-driven evaluation of hadronic contributions
rooted in dispersion theory. 
It is based on sum rules for Compton scattering, of which
a famous example is the  Gerasimov-Drell-Hearn (GDH) 
sum rule \cite{Gerasimov,DH66,Hosoda}:
\beq 
\frac{\al  }{m^2} \varkappa^2   = \frac{1}{2\pi^2}
\int_{\nu_0}^\infty \!\dd \nu \, \frac{\si_{3/2}(\nu)-\sigma_{1/2}(\nu)}{\nu}\,.
\eeq 
On the left-hand side (lhs), we have $\al =e^2/4\pi \simeq 1/137$ the fine-structure constant and $\varkappa$ the AMM of the
spin-1/2 target particle with mass $m$, whereas the 
rhs contains the 
helicity-difference cross section of total photo-absorption
on that particle, integrated over the photon
energy $\nu$, starting from the photo-absorption threshold $\nu_0$.

This
is the sort of relation we are looking for: 
the cross sections can in principle be measured in 
hadronic channels separately (e.g., $\ga \mu \to \pi^0\mu$)
and hence we can  ``measure'' the hadronic contributions to $\varkappa$.
Unfortunately, this strategy would not work here, because
the sum rule involves $\varkappa^2$ and thus we would be probing
a very tiny number  --- recall that the hadronic contribution
to $\varkappa_\upmu$ is of the order  $10^{-8}$.  In powers
of $\al$, 
the hadronic contribution to  $\varkappa_\upmu$ starts at $O(\al^2)$,
therefore the lhs of the GDH sum rule is $O(\al^5)$, whereas
the cross sections of hadronic  photo-production
starts at $O(\al^3)$. This means there is a huge 
(at least 5 orders of magnitude) cancellation under the GDH integral,
and therefore these cross sections would need to be measured
with unprecedented accuracy. 

The same $\varkappa^2$ feature prevents this sum rule from being useful
in theoretical calculations:  to  compute $\varkappa$ to $O(\al^n)$ one needs to know the cross sections
to $O(\al^{2n+1})$, which in fact is a more difficult
calculation. This was explicitly
demonstrated  by Dicus and Vega \cite{DiV01}, who reproduced
the Schwinger's correction ($\al/2\pi$)  through the
GDH sum rule.  
Taking a derivative of the GDH sum rule
with respect to $\varkappa$ linearizes the sum rule
and hence
simplifies the calculations \cite{Pascalutsa:2004ga,Holstein:2005db}. The drawback of the 
GDH-derivative method is that the rhs loses a direct connection to experimental observables:
the helicity-difference cross section is replaced
by a derivative quantity which cannot be accessed in experiment.

Therefore, in what follows we focus on a sum rule which 
is linear in the AMM and involves an observable
cross-section quantity.

\section{The Schwinger sum rule}
\seclab{SSR}
Consider the following
relation, referred to as the Schwinger sum rule
\cite{Schwinger:1975ti,HarunarRashid:1976qz}:
\beq 
\eqlab{newSR}
\varkappa  = \frac{m^2}{\pi^2\alpha}
\int_{\nu_0}^\infty \!\dd \nu \, \left[\frac{\si_{LT}(\nu,Q^2)}{Q}\right]_{Q^2=0}, 
\eeq 
where $\si_{LT}(\nu,Q^2)$ --- the longitudinal-transverse photo-absorption cross section --- is an observable (response
function) corresponding  to an absorption of a 
polarized virtual
photon with energy $\nu$ and space-like 
virtuality $Q^2$ on the target with mass $m$ and 
AMM $\varkappa$, whereby the spin of the target flips.
This response function is rather common
in the studies of nucleon spin structure via electron scattering.
For instance, it plays the central role in the evaluation of the so-called
$\de_{LT}$ polarizability of the proton, and hence in 
the ``$\de_{LT}$ puzzle'' (cf., Ref.~\cite{Hagelstein:2015egb} for a recent review).
One can introduce it for the muon as well, 
and benefit from the fact that 
the sum rule is linear in $\varkappa$, rather than  quadratic.
However, before applying it to the muon case, let us
briefly see how it comes about.

The Schwinger sum rule  can be viewed as
a consequence of the Burkhardt-Cottingham (BC)  and GDH sum rules;
in fact, a linear combination of those.
Introducing the spin-structure functions
$g_{1}(x,Q^2)$
and  $g_{2}(x,Q^2)$ of the spin-1/2 target, with $x=\nicefrac{Q^2}{2m\nu}$  the Bjorken variable,  the BC sum rule
reads as \cite{Burkhardt:1970ti}: $\int_0^1 \dd x \, g_2(x, Q^2)=0$.
Separating the structure functions into the parts
accessed in elastic and inelastic  electron scattering,
the elastic part is expressed in terms of the
Dirac and Pauli form factors, $F_1(Q^2)$ and $F_2(Q^2)$:
\begin{subequations}
\bea 
g_1^{\mathrm{el}}(x,Q^2)&=& \half F_1(Q^2) \big[F_1(Q^2) + F_2(Q^2)]\,\de (1-x) ,\\
g_2^{\mathrm{el}}(x,Q^2)&=& -\mbox{$\frac{Q^2}{8m^2}$} F_2(Q^2) \big[F_1(Q^2) + F_2(Q^2)]
\,\de (1-x) ,\quad
\eea 
\end{subequations}
whereas the inelastic one, $\bar g_i = g_i -g_i^{\mathrm{el}} $, can be expressed
in terms  of the 
response functions $\si_{LT}$ and $\si_{TT}
\equiv 
\nicefrac{1}{2} \,(\si_{1/2} -\si_{3/2})$; for more details see, e.g., Ref.~\cite[Sec.~5.2]{Hagelstein:2015egb}:
\begin{subequations}
\bea 
\bar g_1(x,Q^2)&=& \frac{1}{4\pi^2 \al} \frac{m\nu^3}{\nu^2+Q^2}\left[\frac{Q}{\nu}\si_{LT} + \si_{TT}\right] ,\\
\bar g_2(x,Q^2)&=& \frac{1}{4\pi^2 \al} \frac{m\nu^3}{\nu^2+Q^2}\left[\frac{\nu}{Q}\si_{LT} - \si_{TT}\right].
\eea 
\end{subequations}
In the limit of $Q^2\to 0$, with\footnote{Here the explicit use of $F_1(0)=1$ 
limits the applicability of the resulting sum rule in \Eqref{newSR} to charged particles, in contrast to, e.g., the GDH sum rule which 
holds as well for a neutral particle, such as the neutron. } $F_1(0)=1$
and $F_2(0)=\varkappa$, the BC sum rule
yields:
\bea 
 (1+\varkappa)\varkappa &=& \lim_{Q^2\to 0} \frac{8m^2}{Q^2}
 \int_0^{x_0} \dd x \,  \bar
 g_2(x,Q^2),\nn\\
 &=& \frac{m^2}{\pi^2\al}\int_{\nu_0}^\infty \dd\nu \left[ 
 \frac{\si_{LT}}{Q} - \frac{\si_{TT}}{\nu}\right]_{Q^2=0},
 \eea 
where $x_0 = \nicefrac{Q^2}{2m\nu_0}$ is the inelastic threshold of the Bjorken variable.
Now, the GDH sum rule allows us to cancel the $\varkappa^2$ on the lhs 
against the $\si_{TT}$ term on the rhs, resulting in \Eqref{newSR}.
The latter can also be rewritten in terms of the spin structure functions as:
\beq 
\varkappa = \lim_{Q^2\to 0} \frac{8m^2}{Q^2}
 \int_0^{x_0} \dd x \;  
 [\bar g_1+\bar g_2](x,Q^2).
\eeq 
Thus, we ``only'' need to know how (a moment of) the muon spin-structure function combination $g_1+g_2$ is affected by hadronic contributions.

\section{Hadronic contributions via the sum rule}

Let us now examine the hadronic contributions to $(g-2)_\upmu$ using the Schwinger sum rule.
The first thing to consider is the hadron production on the muon
shown in \Figref{Hphotoprod}, i.e., $\ga \mu \to \mu + \mathrm{hadrons}$. Examples of these processes are:
$\ga \mu^\pm \to \pi^0\mu^\pm$, 
$\ga \mu^\pm \to \pi^+\pi^-\mu^\pm$. 
Here one can distinguish two mechanisms, 
timelike Compton scattering [\Figref{Hphotoprod}(a)], and 
the Primakoff effect [\Figref{Hphotoprod}(b)]. They add up
incoherently (i.e., there is no interference term) because
of $C$-parity conservation, {\it viz.}, Furry's theorem.
\begin{figure}[htb]
\includegraphics[width=0.37\textwidth]{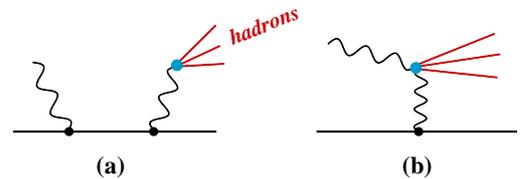}
\caption{Mechanisms
of hadron photo-production off a lepton: (a) timelike Compton scattering (crossed diagram omitted), (b) Primakoff effect.}
\figlab{Hphotoprod}
\end{figure}

Both mechanisms begin to contribute at $O(\al^3)$ to $\sigma_{LT}$
and hence at  $O(\al^2)$ to $g-2$. The first mechanism 
(timelike CS) corresponds with the HVP contribution
[\Figref{Hamm}(a)], and thus provides an alternative access to it,
see \secref{HVP}. On the other hand, the leading-order 
contribution of the Primakoff mechanism 
to $g-2$ should vanish exactly, as it does not correspond with the
HVP contribution, and there is no other hadronic contribution
to $g-2$ at this order. An explicit proof of this statement 
[i.e., vanishing effect of \Figref{Hphotoprod}(b)
on $g-2$]
should be possible through the use of the light-by-light scattering sum rules \cite{Pascalutsa:2010sj,Pascalutsa:2012pr}. As a result, the Primakoff mechanism can only contribute in interference
with subleading effects, such as the one shown in \Figref{pi02gamma}
for the case of $\pi^0\gamma$ and $\pi^0$ production.

The main advantage of using the Schwinger sum rule, however, 
is that one need not be concerned
with computing the subleading effects of hadronic production --- 
they all can in principle be measured experimentally. This can be achieved at an electron-muon collider with polarized beams needed to access the spin structure
functions. Tagging is not necessary, since we only need the quasi-real-photon limit. No separation of radiative corrections is necessary:
as long as hadrons are present in the final state, they are part
of the hadronic contributions to the spin structure functions of one of the leptons. In fixed-target experiments, one would need to measure the recoil electron polarization.

\begin{figure}[tb]
\includegraphics[width=0.3\textwidth]{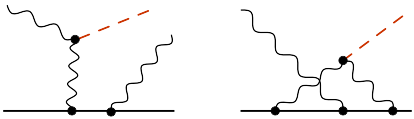}
\caption{Subleading mechanisms
accompanying the single-meson photo-production.}
\figlab{pi02gamma}
\end{figure}

\begin{figure}[htb]
\includegraphics[width=0.13\textwidth]{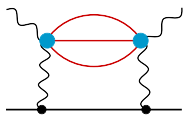}
\caption{The HLbL contribution to Compton scattering.}
\figlab{HCompton}
\end{figure}
\begin{figure}[b]
\includegraphics[width=0.32\textwidth]{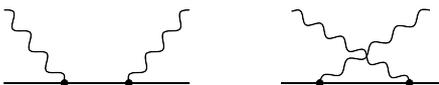}
\caption{Tree-level Compton scattering diagrams.}
\figlab{treeCS}
\end{figure}

Apart from the abovementioned hadron-production channels, the muon structure functions can be affected by hadrons in the loops. The most important (in orders of $\al$)  is
the effect of HLbL on the Compton scattering (CS),
shown in \Figref{HCompton}, interfering with the tree-level Compton effect, \Figref{treeCS}. 
Note that here the initial photon
is quasi-real, whereas the final one is real. Thus, the evaluation of the HLbL contribution to $\si_{LT}$ involves the HLbL amplitude with only two virtual photons. This is substantially simpler 
than the corresponding HLbL contribution to $g-2$ shown in 
\Figref{Hamm}(b), which involves the LbL amplitude with three
virtual photons and one quasi-real.

\begin{figure}[pt]
\includegraphics[width=0.32\textwidth]{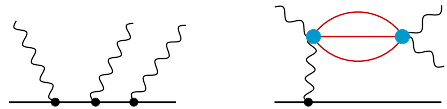}
\caption{The HLbL contribution through two-photon production.}
\figlab{Hmulti}
\end{figure}

Another HLbL effect, of the same order in $\al$, arises from the interference of the diagrams in \Figref{Hmulti}, describing the hadronic
contribution to the $\mu \ga\ga $ channel. Here the treatment of the HLbL contribution is even simpler than in the Compton channel, since the HLbL amplitude is not in the loop and only one of the four photons is virtual.

Before considering these hadronic contributions further, 
it is instructive to compute the leading QED contribution of \Figref{treeCS} by itself.
A straightforward calculation yields:\footnote{This cross section is not only given by the interference of the two diagrams in \Figref{treeCS}, as one could have naively expected by comparing the topology of the resulting contributions to the forward doubly-virtual CS amplitude. The
influence of the one-particle irreducible graphs in the latter amplitude is negated by the integration over $\nu$ in the sum rule.}
\bea  
&& \si^{\ga \mu \to \ga\mu }_{LT}(\nu, Q^2) =\frac{\pi  \alpha ^2 Q \,(s-m^2)^2 }{4 m^3 \nu ^2 \big(\nu ^2+Q^2\big)}\\
&&\times \left( -2 - \frac{m(m+\nu)}{s}
+ \frac{3 m+2 \nu}{\sqrt{\nu ^2+Q^2}}
   \, \mathrm{arccoth} \frac{m+\nu}{\sqrt{\nu ^2+Q^2}}\right),\nn
\eea
with $s=m^2+2m\nu-Q^2$. 
Substituting this expression into the Schwinger sum rule, one obtains
for $\varkappa$ the Schwinger correction: $\al/2\pi$.
This exercise thus provides a check of the sum rule in
leading-order QED, similar to the one done by
Tsai et al.~\cite{Tsai:1975tj}.

\subsection{Hadronic vacuum polarization}
\seclab{HVP}
To reproduce the
leading HVP contribution [\Figref{Hamm}(a)] through the
hadron photo-production mechanism shown in \Figref{Hphotoprod}(a),
we factorize the invariant mass distribution $\dd\si(\ga \mu \to  \mu X )/\dd M_X^2$, arising from 
\Figref{Hphotoprod}(a), into the cross sections of timelike Compton scattering
$\si(\ga \mu \to \ga^\ast \mu)$ and of the subsequent photon decay into
hadrons $\Gamma (\ga^\ast\to X)$, cf. Appendix~\ref{App1}. The latter cross section is, by unitarity, expressed via the absorptive part of the hadronic contribution to vacuum polarization,
$\im \Pi^{\mathrm{had}}(q^{\prime\,2})$. 
The tree-level $LT$ cross section of Compton scattering, with initial and final photon virtualities respectively 
given by $q^2=-Q^2$ and $q^{\prime \,2} =M^2_{X}$, is easily computed 
to yield:
\bea  
&& \left[\frac{\si^{\ga \mu \to \ga^\ast\mu }_{LT}(\nu,Q^2)}{Q}\right]_{Q^2= 0}  =  
\frac{\pi  \alpha ^2  }{2 m^2\nu^3 }   \\
&& \qquad \times\, \Big[-(5s+m^2+M_X^2) \la +
(s+2m^2-2M_X^2)  \log\frac{\beta+\la}{\beta-\la} \Big], \nn
\eea 
with $s=m^2+2m\nu$, $\be = (s+m^2-M_X^2)/2s$, 
and $\la =(1/2s)\sqrt{[s-(m+M_X)^2]\, [s-(m-M_X)^2]} $.
From the Schwinger sum rule we then have:
\begin{widetext}
\bea  
\eqlab{SRforHVP}
\varkappa &= &\frac{m^2}{\pi^2\alpha}
\int\limits_{4m_\pi^2}^\infty \dd M_X^2 \int\limits_{\nu_0}^\infty \!\dd \nu \, 
 \left[\frac{1}{Q}\frac{\dd \si^{\ga \mu \to \mu X }_{LT}(\nu,Q^2)}{\dd M^2_X}\right]_{Q^2= 0} = \frac{1}{\pi}\int\limits_{4m_\pi^2}^\infty \dd M_X^2 \, \frac{\im \Pi^{\mathrm{had}}(M_X^{2})}{M_X^2}
\frac{m^2}{\pi^2\alpha}
\int\limits_{\nu_0}^\infty \!\dd \nu \, \left[\frac{\si^{\ga \mu \to \ga^\ast\mu }_{LT}(\nu,Q^2)}{Q}\right]_{Q^2= 0}
\eea 
where $\nu_0= M_X (1+M_X^2/2m^2)$ is the 
photo-production threshold, while $4 m_\pi^2$ is set
by the lightest produced state (here, $\pi^-\pi^+$ pair).
Finally, performing the integration over $\nu$, we obtain: 
\bea  
\eqlab{SRinMQED}
&&\frac{m^2}{\pi^2\alpha}
\int_{\nu_0}^\infty \!\dd \nu \, \left[\frac{\si^{\ga \mu \to \ga^\ast\mu }_{LT}(\nu,Q^2)}{Q}\right]_{Q^2= 0} \, =\,  \frac{\al}{\pi} K(M_X^2/m^2)  \, \equiv\,  \frac{\al}{\pi} \int_0^1 \dd x \frac{x^2(1-x)}{x^2 + (1-x)(M_X^2/m^2)} , 
\eea  
\end{widetext}
and hence the standard expression for the HVP contribution (see, e.g.,
Ref.~\cite{Jegerlehner:2017gek}),
\beq 
\varkappa^{\mathrm{HVP}} = \frac{\al}{\pi^2}\int_{4m_\pi^2}^\infty \dd s\, 
K(s/m^2) \, \frac{\im \Pi^{\mathrm{had}}(s)}{s},
\eeq 
is exactly reproduced.

In practice, a determination of the HVP contribution through the
Schwinger sum rule has an important conceptual difference 
from the standard
practice of measuring $e^+e^-\to \mathrm{hadrons}$. The latter method involves an approximation
of the single-photon exchange, the two-photon exchange effects ought to
be removed. In the sum-rule method, the two-photon-exchange and
other subleading effects need not be removed, they are part of the
sought hadronic contribution.


Further novel features of calculating the hadronic 
contributions through the Schwinger sum rule can be seen
in the following example of the meson-exchange contribution.

\subsection{ Pseudoscalar meson  contribution}
\seclab{pion}
The neutral pseudoscalar mesons $\pi^0$ and $\eta$ play a significant 
role in the HLbL contribution through the mechanism shown
in \Figref{HLbLpi0}. Let us see how this mechanism is evaluated
using the Schwinger sum rule.  
\begin{figure}[b]
\includegraphics[width=0.16\textwidth]{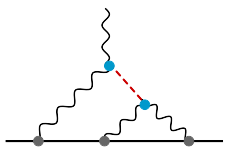}
\caption{$\pi^0$-exchange HLbL contribution to $(g-2)_\upmu$.}
\figlab{HLbLpi0}
\end{figure}

In the hadronic channel, we need to know the
$LT$ cross section for the single-meson photo-production
off the lepton. This can in principle be measured directly,  or calculated to leading-order evaluating the diagrams in
\Figref{Hphotoprod2}. Note that, in addition to the Primakoff mechanism (last diagram), we have here the subleading (in $\al$) mechanisms of
the type given by the second diagram of \Figref{pi02gamma}. 
The latter is effectively accounted for in the first two
graphs in \Figref{Hphotoprod2}, where the meson-lepton-lepton ($\pi \ell \ell$) coupling is fixed from the decay width of pseudoscalar mesons into leptons
(i.e., $\pi^0\to e^+ e^-$ and $\eta\to \mu^+ \mu^-$).

\begin{figure}[tb]
\includegraphics[width=0.42\textwidth]{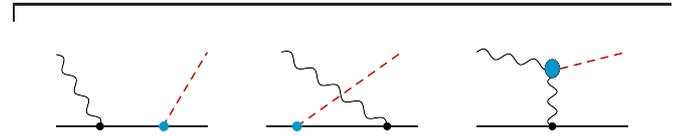}
\caption{Single-meson photo-production off a lepton.}
\figlab{Hphotoprod2}
\end{figure}

The same experimental information on pseudoscalar-into-leptons decays fixes the counter-term needed to renormalize
the vertex calculation of the form factor in \Figref{vertex}, which is
needed in the Compton channel calculation, cf.~\Figref{pi0CS}(a).
It is interesting that this $\pi \ell \ell$ form factor
enters profoundly in the calculation of $\pi^0$ exchange
in the  hyperfine splitting of muonic hydrogen \cite{Hagelstein:2015lph,Huong:2015naj,Dorokhov2017,Hagelstein:2017cbl}.
The $\pi^0$ effects in $(g-2)_\upmu$ and muonic hydrogen are thus
interrelated.

\begin{figure}[tb]
\includegraphics[width=0.42\textwidth]{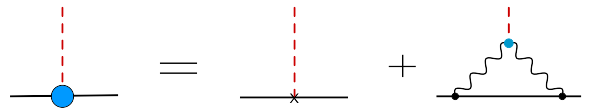}
\caption{Pseudoscalar meson coupling to leptons.}
\figlab{vertex}
\end{figure}

\begin{figure}[tb]
\includegraphics[width=0.47\textwidth]{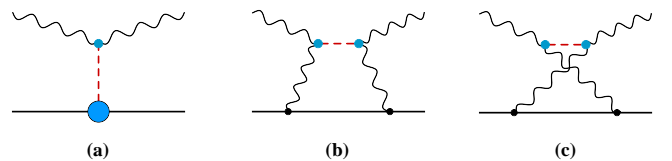}
\caption{Pseudoscalar-meson contribution to Compton scattering (a) $t$-channel exchange with the $\pi \ell \ell$ form factor given by 
\Figref{vertex}; (b) and (c) show the direct and crossed box, respectively. Drawn using
Jaxodraw \cite{Binosi:2003yf}.}
\figlab{pi0CS}
\end{figure}

Furthermore, there are contributions from the $2\ga$ and  $\pi^0 \gamma$ channels given respectively by \Figref{Hmulti} with $\pi^0$ as the virtual hadronic state and the first diagram of \Figref{pi02gamma}. 

It is important to realize that calculating the $\pi^0$
contribution through the diagrams in Figs.~\ref{fig:Hphotoprod2} and \ref{fig:pi0CS} and the multi-particle channels, we do not need the  transition
form factor (TFF) with two virtual photons. We only need 
the TFF for a single virtual photon [i.e, 
$F_{\pi^0 \ga\ga^\ast}(Q^2)$] in the box graphs of 
\Figref{pi0CS}(b) and (c), since the external photons are (quasi-)real. 
The doubly-virtual TFFs could be used in evaluation
of the $\pi\ell\ell$ form factor \Figref{vertex}. However,
their impact therein is largely diminished by the renormalization
and the use of the empirical width.

\section{Conclusion}
We have considered the hadronic contributions (HVP and HLbL) to 
$g-2$, and showed how they can be assessed using
the Schwinger sum rule presented in \secref{SSR}. The sum rule
separates the hadronic contributions into two types: 
\begin{itemize}
\item[(i)]
hadron photo-production (\Figref{Hphotoprod}), 
\item[(ii)]
HLbL contribution to non-hadronic channels (\Figref{HCompton}, \ref{fig:Hmulti}). 
\end{itemize}
Type (i)  has a clear relation to observables. These
are the spin structure functions of the lepton with hadrons
in the final state (hadronic channels) that
could in principle be measured in electron-muon
collisions. This type of 
contributions is  readily suited for a 
model-independent, ``data-driven'' evaluation. 

In type (ii), the hadrons may appear in the loops, similar to the
sought HLbL contribution to $g-2$ [\Figref{Hamm}(b)].
However, the sum rule evaluation 
only requires the HLbL for the situation of two virtual photons
forming a loop, rather than three virtual photons forming two loops as required in the direct evaluation of \Figref{Hamm}(b).
The former evaluation, therefore, requires much less information about HLbL, and is technically simpler. For example, 
the evaluation of the neutral-pion, and other single-meson
contributions, will only require the transition form factor 
to one real and one virtual photon ($M\to \ga\ga^\ast$),
rather than two virtual photons ($M\to \gamma^\ast\gamma^\ast$) as
required usually.

The Schwinger sum rule is thus a very promising tool for
a data-driven evaluation of the hadronic contributions
to $(g-2)_\upmu$. Despite being quite different from the
existing dispersive approaches, the present approach
may benefit from 
the dispersive analysis of the HLbL amplitude by Colangelo {\it et al.}~\cite{Colangelo:2014pva,Colangelo:2014dfa,Colangelo:2017qdm,Colangelo:2017fiz}, trimmed to the narrower kinematical range required for the evaluation of \Figref{HCompton}. In a more distant perspective,
the type (ii) contribution will be calculable in lattice QCD.

In \secref{HVP}, we have reproduced the standard expression for the HVP contribution via the sum rule. In \secref{pion}, we have outlined how the sum rule program works for
the pseudoscalar meson contributions. With very few modifications
it applies, of course, to the axion contributions to $g-2$, which have
lately been receiving renewed attention in connection
with collider searches \cite{Marciano:2016yhf,Bauer:2017nlg}.

The advantages of evaluating the axion and other beyond-SM 
contributions by using the Schwinger sum rule are less obvious
than in the hadronic case, where data-driven approaches are generally desirable in the absence of precise {\it ab initio} calculations. And, even in a more advanced lattice-QCD era,
the presented sum-rule approach may be
advantageous, if only for its
clear-cut separation of the explicit
hadron production from the virtual hadronic effects.

\acknowledgments

This work was supported by the Deutsche Forschungsgemeinschaft (DFG) 
through the Collaborative Research Center [The Low-Energy Frontier of the Standard Model (SFB 1044)] and in part by the Swiss National Science Foundation. 
\appendix
\begin{widetext}

\section{Factorizing the hadron photo-production through timelike Compton scattering}\label{App1}
 Here we show how the hadron photo-production process $(\ga \mu \to  \mu X )$, going through the
 mechanism in \Figref{Hphotoprod}(a), can be decomposed into
the timelike Compton scattering $(\ga \mu \to \ga^\ast \mu)$ and the virtual-photon decay into
hadrons $(\ga^\ast\to X)$. The factorization applies to all helicity cross sections, including $\si_{LT}$. Denoting the incoming (outgoing) muon and photon 4-momenta as $p$ ($p'$) and $q$ ($q'$), and 
the 4-momenta of the particles in $X$ as $k_i$,
the total photo-absorption cross section is given by:
\beq
\si(\ga \mu \to  \mu X )=\frac{(2\pi)^4}{4I}\int\dd^4q' \int\prod_i\frac{\dd^3 \boldsymbol{k}_i}{2E_{k_i}(2\pi)^3} \int \frac{\dd^3 \boldsymbol{p'}}{2E_{p'}(2\pi)^3}\left[\frac{\Lambda^{\dagger\mu}\Lambda^{\nu}\rho_{\mu\nu}}{(-q^{\prime\,2})^2}\right]  \delta^4(q'-\sum_i k_i)\,\delta^4(p+q-p'-q'),\eqlab{CS1}
\eeq
with $I^2=(p\cdot q)^2-p^2\,q^2$ the initial flux factor,
$\Lambda^\mu$ the virtual-photon decay vertex, and $\rho^{\mu \nu}$  the squared matrix element of the timelike Compton scattering, where the vector indices refer to the virtual photon. The integration over $q'$ 
is conveniently introduced at the expense of inserting a compensating $\de$-function. The other integrations cover the phase space of the final state to form a total cross section.

The photon decay width is defined as:
\begin{subequations}
\eqlab{A2}
\bea
\left[\Gamma (\ga^\ast\to X)\right]^{\mu \nu}&=&\int \prod_i\frac{\dd^3 \boldsymbol{k_i}}{2E_{k_i}(2\pi)^3}\frac{\Lambda^{\dagger\mu}\Lambda^{\nu}}{2E_{q'}} (2\pi)^4 \delta^4(q'-\sum_i k_i), \\ &=&-\frac{1}{\sqrt{q^{\prime\,2}}}\,(q^{\prime\,2} g^{\mu \nu}-q'^\mu q'^\nu) \im \Pi_{X} (q^{\prime\,2}),\eqlab{CS2}
\eea
\end{subequations}
where in the last step we made use of its transverse tensor structure and unitarity, with $\Pi_{X}$ being the contribution of state $X$ to the vacuum polarization. 


Substituting \Eqref{A2} into \Eqref{CS1}, and using gauge invariance ($q'^\mu \rho_{\mu \nu}=0$) to drop the $q'^\mu q'^\nu$ term in \Eqref{CS2}, we find:
\beq
\si(\ga \mu \to  \mu X )=-\frac{1}{2I}\int\dd^4q' \int \frac{\dd^3 \boldsymbol{p'}}{2E_{p'}(2\pi)^3}\,\rho^\mu_\mu\,\frac{\im \Pi_{X}(q^{\prime\,2})}{q^{\prime\,2}}\, \delta^4(p+q-p'-q').
\eeq
Now we can identify the total cross section of the
timelike Compton scattering, $\sigma(\ga \mu \to \ga^\ast \mu)$, and thus arrive at:
\beq
\si(\ga \mu \to  \mu X )=\frac{1}{\pi}\int\frac{\dd M_X^2}{M_X^2}\,\sigma(\ga \mu \to \ga^\ast \mu) \im \Pi_{X}(M_X^2).
\eeq
The remaining integral over $M_X^2 = q^{\prime\,2}$ reflects
the fact that we are integrating over all possible states $X$.
\end{widetext}

\end{document}